\documentclass{nature}
\usepackage{graphicx}
\usepackage{caption}
\usepackage{amsmath}
\usepackage{textcomp}
\usepackage{url}
\usepackage{lineno}
\usepackage{gensymb}
\usepackage[font=singlespacing]{caption}
\usepackage{siunitx}
\usepackage{upgreek}
\usepackage{textgreek}
\usepackage{pdfpages}

\makeatletter
\let\saved@includegraphics\includegraphics
\AtBeginDocument{\let\includegraphics\saved@includegraphics}
\renewenvironment*{figure}{\@float{figure}}{\end@float}
\makeatother

\makeatletter
\long\def\@makecaption#1#2{%
  \vskip\abovecaptionskip
  \sbox\@tempboxa{#1 #2}%
  \renewcommand{\baselinestretch}{0.9}\reset@font
  \ifdim \wd\@tempboxa >\hsize
    #1 #2\par
  \else
    \global \@minipagefalse
    \hb@xt@\hsize{\hfil\box\@tempboxa\hfil}%
  \fi
  \vskip\belowcaptionskip}
\makeatother

\begin{document}
\includepdf[page={1,2,3,4,5,6,7,8,9,10,11,12,13,14,15,16}]{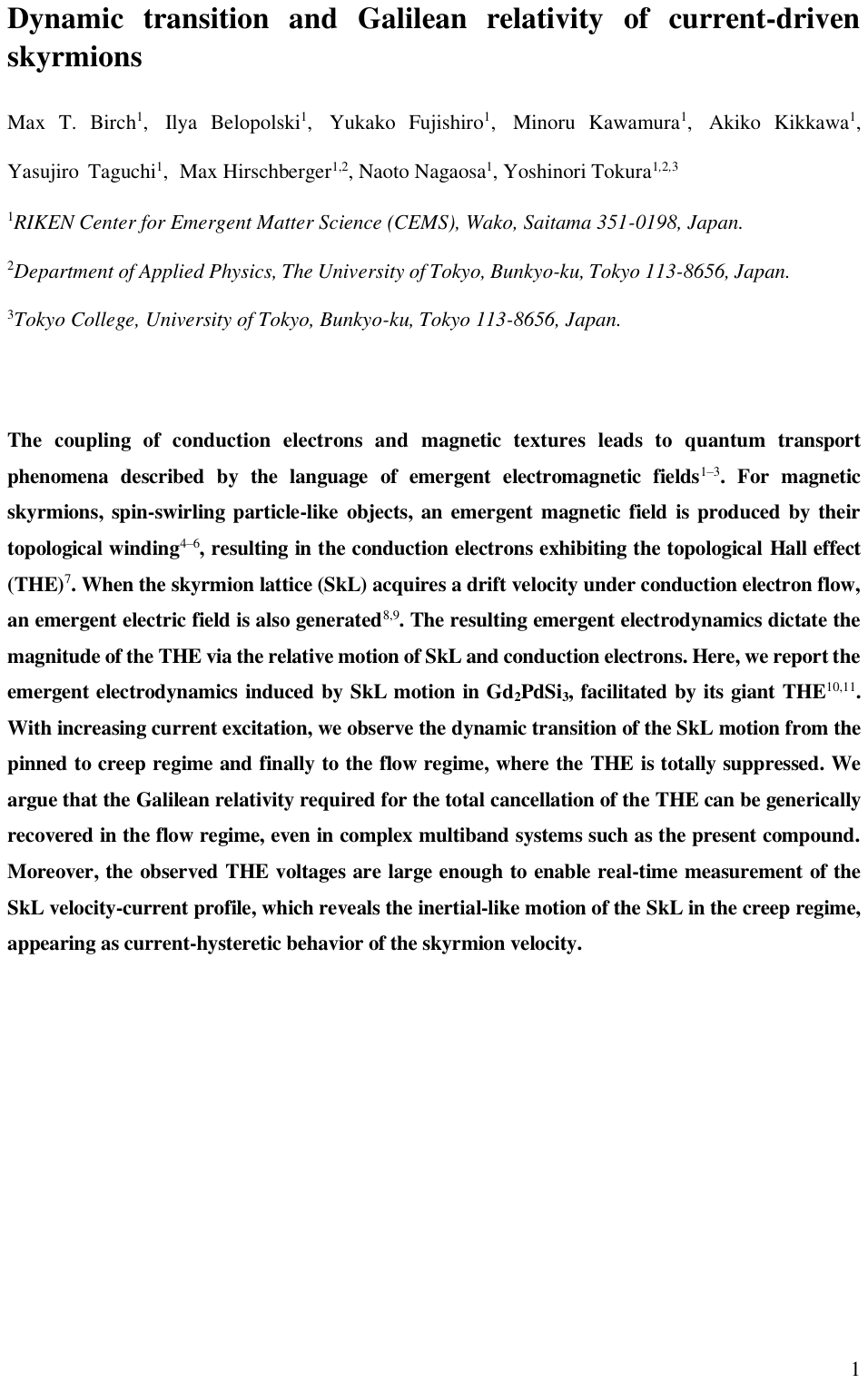}


\renewcommand\thefigure{\arabic{figure}}    
\renewcommand\figurename{Extended Data Figure }
\makeatletter
\def\fnum@figure{\figurename\thefigure}
\makeatother
\setcounter{figure}{0}   

\begin{figure}
\centering
\includegraphics[width=0.7\textwidth]{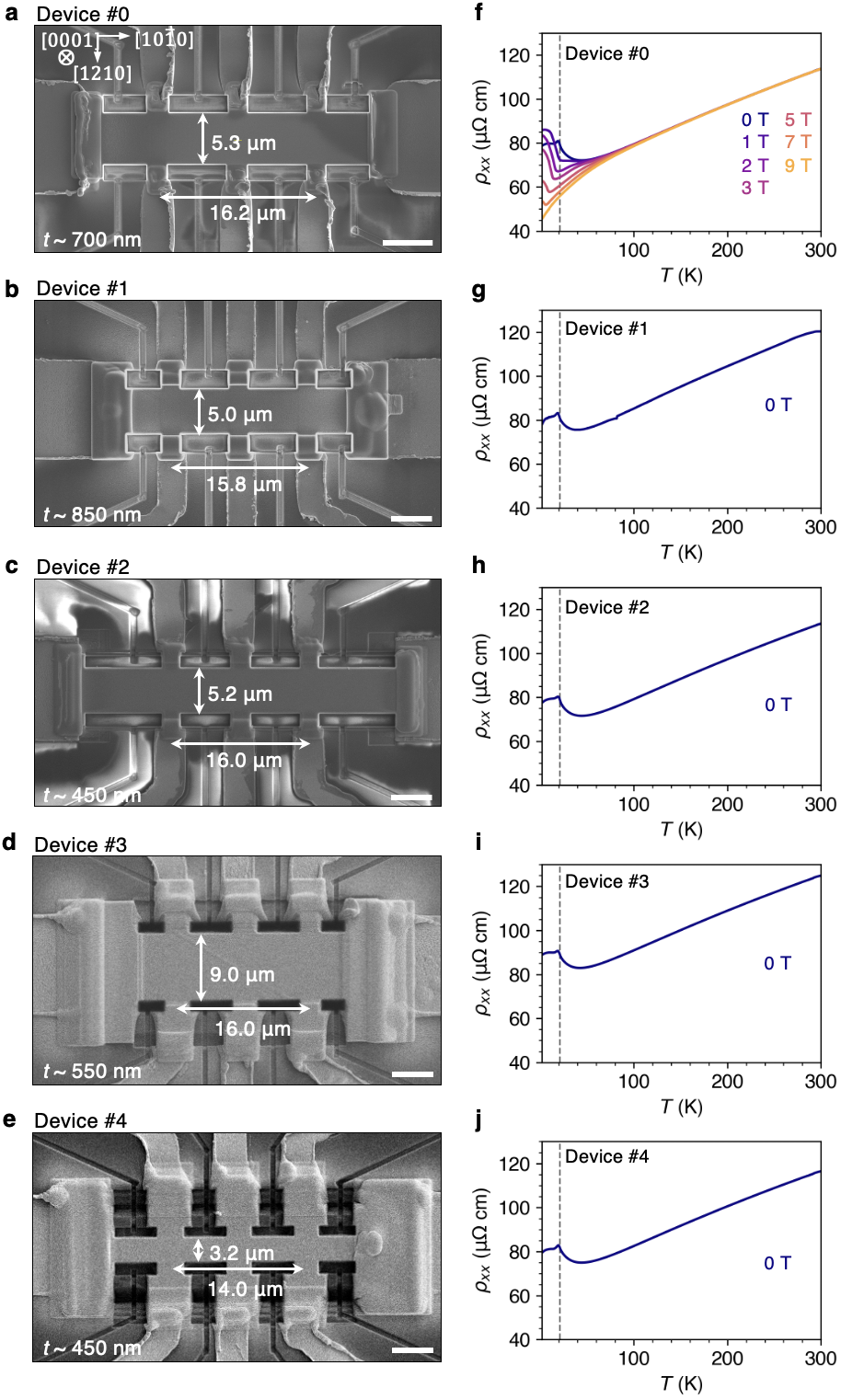}
\caption{$\vert$ \textbf{Characterisation of the focused ion beam-fabricated Gd$_2$PdSi$_3$ devices.} \textbf{a}-\textbf{e} Scanning electron microscopy images of the fabricated Gd$_2$PdSi$_3$ devices. Relevant sample dimensions are labelled, and the scale bar is 5 \textmu m. \textbf{f}-\textbf{j} Longitudinal resistivity $\rho_{xx}$ measured as a function of decreasing temperature $T$ at a selection of applied magnetic fields, for Devices \#0 to \#4, respectively.
}
\label{fig_E1}
\end{figure}

\begin{figure}
\centering
\includegraphics[width=1\textwidth]{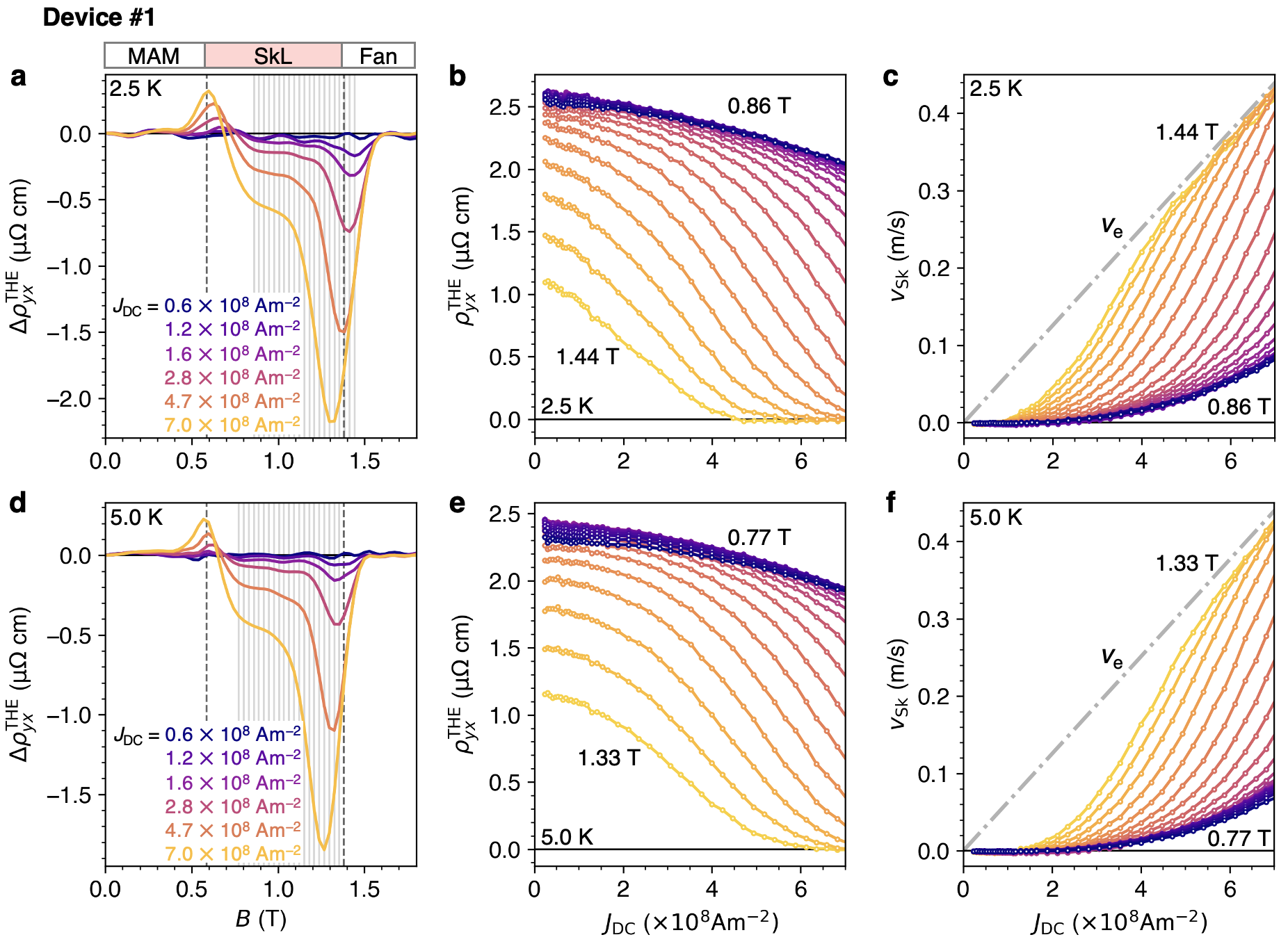}
\caption{$\vert$ \textbf{Additional skyrmion velocity measurements via the nonlinear topological Hall effect -- Device \#1.} \textbf{a} The change in the measured Hall resistivity, $\rho_{yx}^{J\rightarrow 0}-\rho_{yx}^{J}$, at 2.5~K measured as a function of the applied magnetic field $B$ at different DC current densities, $J_{\rm{DC}}$, for Device \#1. The boundaries between the ground state magnetic phases are labelled by the grey dashed lines. The vertical solid grey lines highlight fields considered in later panels. \textbf{b} The topological Hall resistivity plotted as a function of $J_{\rm{DC}}$ for selected $B$ between 0.86 and 1.44 T. \textbf{c} The calculated skyrmion velocity, $v_{\rm{Sk}}$, plotted as a function of $J_{\rm{DC}}$ for selected $B$ between 0.86 and 1.44 T. The electron drift velocity, $v_{\rm{e}}$, is plotted as the dotted/dashed line, calculated from the measured charge carrier density. \textbf{d}-\textbf{f} The same as \textbf{a}-\textbf{c}, but at 5.0 K.
}
\label{fig_E2}
\end{figure}

\begin{figure}
\centering
\includegraphics[width=1\textwidth]{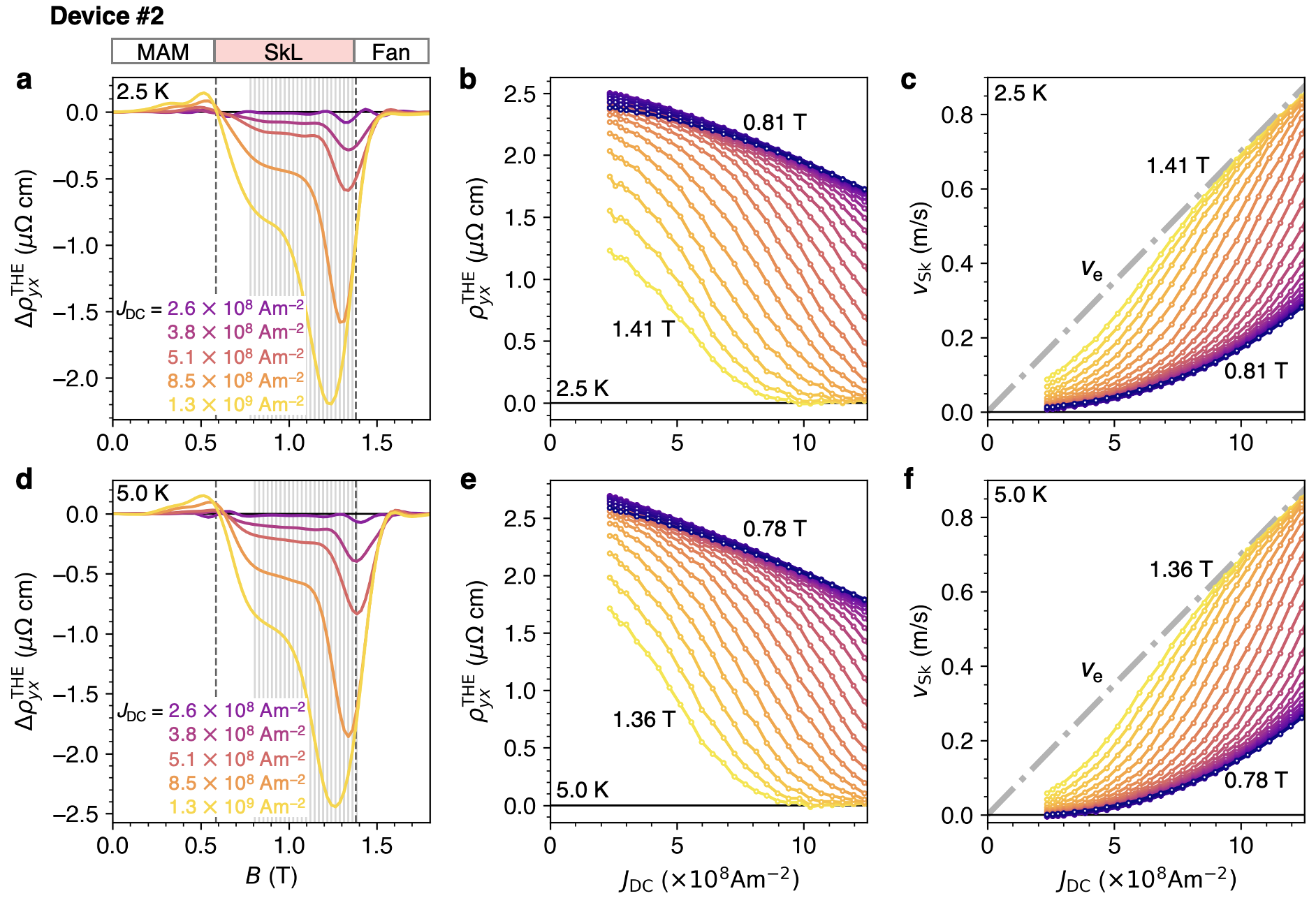}
\caption{$\vert$ \textbf{Additional skyrmion velocity measurements via the nonlinear topological Hall effect -- Device \#2} \textbf{a} The change in the measured Hall resistivity, $\rho_{yx}^{J\rightarrow 0}-\rho_{yx}^{J}$, at 2.5~K measured as a function of the applied magnetic field $B$ at different DC current densities, $J_{\rm{DC}}$, for Device \#1. The boundaries between the ground state magnetic phases are labelled by the grey dashed lines. The vertical solid grey lines highlight fields considered in later panels. \textbf{b} The topological Hall resistivity plotted as a function of $J_{\rm{DC}}$ for selected $B$ between 0.86 and 1.44 T. \textbf{c} The calculated skyrmion velocity, $v_{\rm{Sk}}$, plotted as a function of $J_{\rm{DC}}$ for selected $B$ between 0.86 and 1.44 T. The electron drift velocity, $v_{\rm{e}}$, is plotted as the dotted/dashed line, calculated from the measured charge carrier density. \textbf{d}-\textbf{f} The same as \textbf{a}-\textbf{c}, but at 5.0 K.
}
\label{fig_E3}
\end{figure}

\begin{figure}
\centering
\includegraphics[width=1\textwidth]{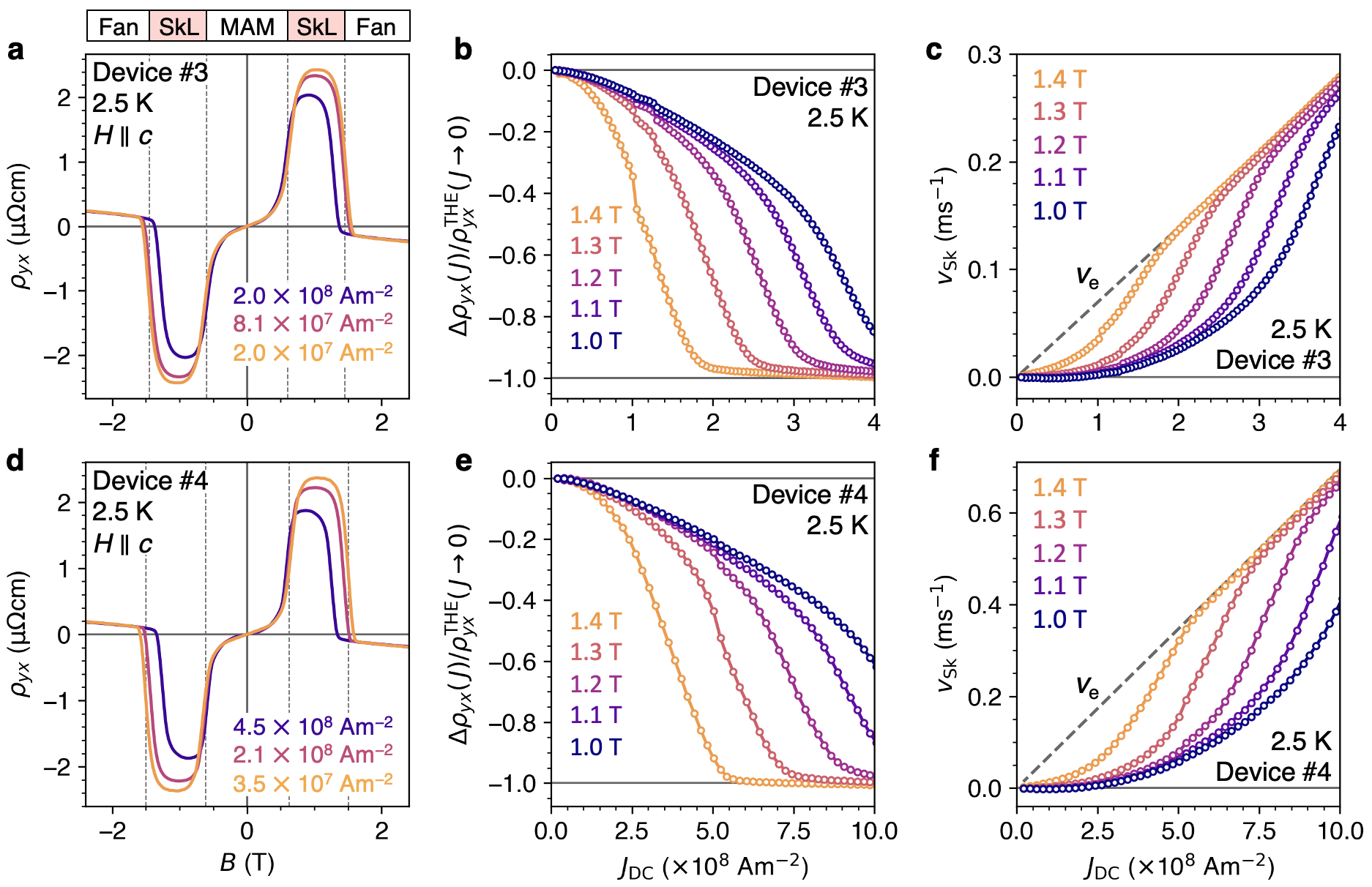}
\caption{$\vert$ \textbf{Additional skyrmion velocity measurements via the nonlinear topological Hall effect -- Device \#3 and \#4.} \textbf{a} The Hall resistivity $\rho_{yx}$ measured at 2.5~K as a function of the applied magnetic field $B$, at three current densities $J_{DC}$ for Device \#3. \textbf{b} The change in the Hall resistivity $\Delta \rho_{yx}=\rho_{yx}(J)-\rho_{yx} (J\rightarrow0)$, normalized by the zero current limit value of the Hall resistivity, $\rho_{yx} (J\rightarrow0)$, measured as a function of $J_{DC}$ at 2.5 K and various applied fields in Device \#3. \textbf{c}, The skyrmion velocity, $v_{\rm{Sk}}$ plotted as a function of $J$, calculated from the data in \textbf{b}, for various applied fields in Device \#3. \textbf{d}-\textbf{f} The same, but for Device \#4.
}
\end{figure}

\begin{figure}
\centering
\includegraphics[width=0.5\textwidth]{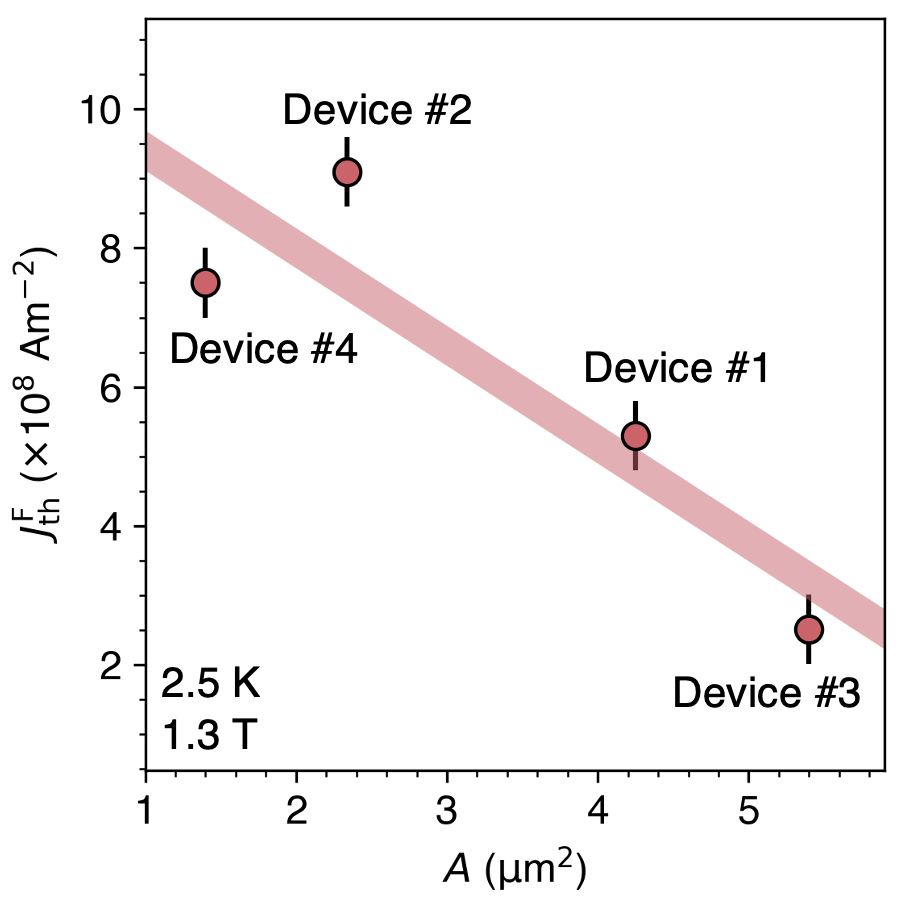}
\caption{$\vert$ \textbf{Sample dependence of the current density threshold to enter the flow regime.} The estimated current density threshold to enter the flow regime, $J_{\rm{th}}^{\rm{F}}$, at an applied field of 1.3 T and sample temperature of 2.5 K, plotted as a function of the cross sectional area $A$ of each FIB Device \#1-4. Error bars show the standard error. An approximate fitted linear trend is plotted as the solid line. 
}
\label{fig_E5}
\end{figure}

\begin{figure}
\centering
\includegraphics[width=0.87\textwidth]{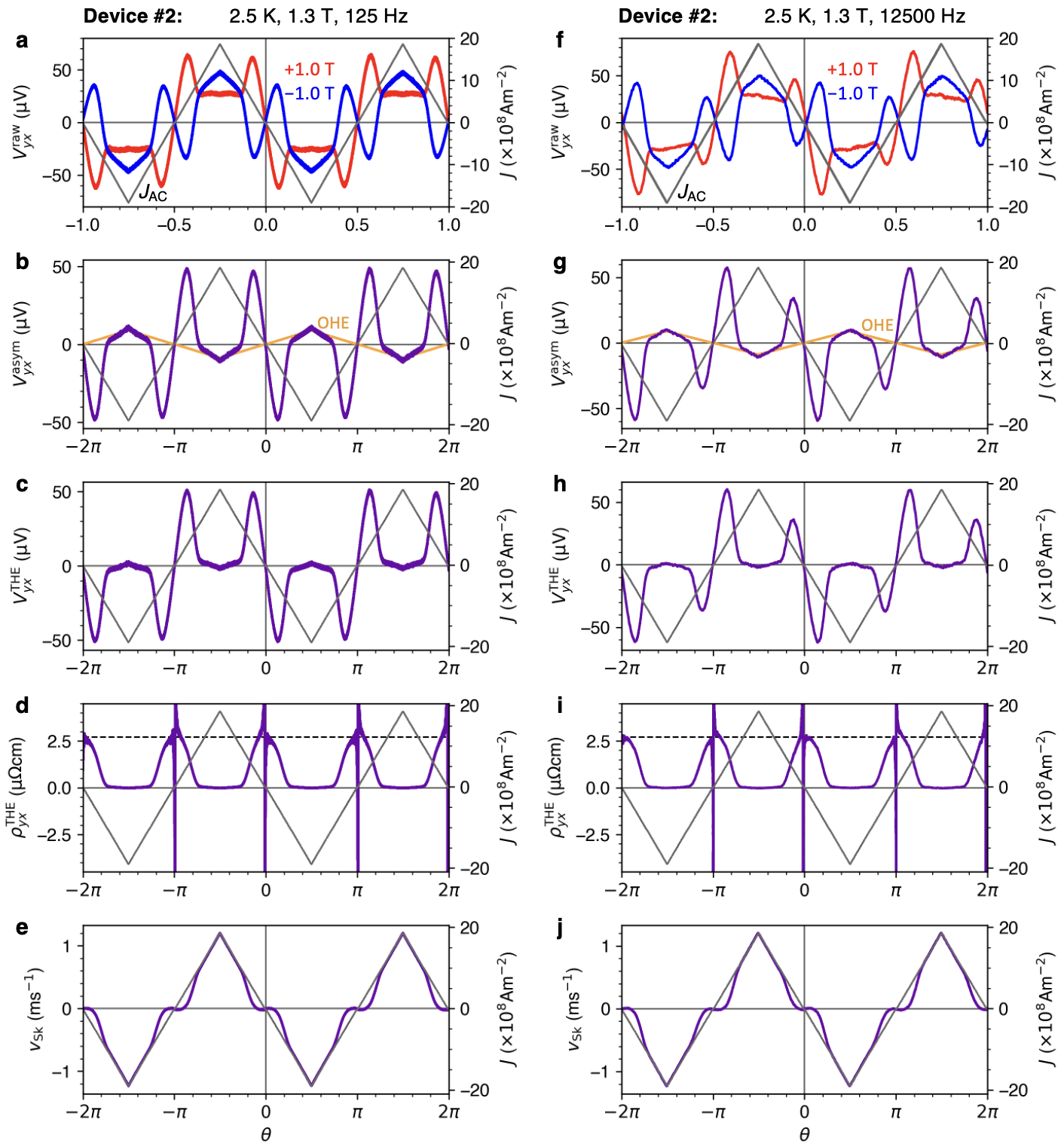}
\caption{$\vert$ \textbf{Analysis of the oscilloscope Hall measurements.} In all panels, the solid grey line plots the applied current density $J$, with a peak $J_{\rm{AC}}^{\rm{max}}$ of 1.9$\times$10$^{9}$ Am$^{-2}$. \textbf{a} The measured Hall voltage response $V_{yx}^{\rm{raw}}$ of Device \#2, measured at 2.5~K and with an AC current at 125~Hz (corresponding to a $dJ/dt$ = 1 $\times$ 10$^{12}$ Am$^{-2}$s$^{-1}$), plotted as a function of $\theta$. Data was acquired at both $\pm$ 1.0 T, displayed as the red and blue lines respectively. \textbf{b} The voltage traces acquired at $\pm$ 1.0 T were antisymmetrised, yielding $V_{yx}^{\rm{asym}}$ (purple line). The estimated contribution from the ordinary Hall effect (OHE) is plotted (orange line). The contribution from the topological Hall effect (THE) was acquired by subtracting this estimated OHE signal. \textbf{c}, The calculated topological Hall effect resistivity $\rho_{yx}^{\rm{THE}}$ plotted as a function of $\theta$. \textbf{d} The calculated skyrmion velocity $v_{\rm{Sk}}$ plotted as a function of $\theta$. \textbf{e}-\textbf{h} The same as \textbf{a}-\textbf{d}, but measured with a current density oscillating at a frequency of 12500~Hz (corresponding to a $dJ/dt$ = 1 $\times$ 10$^{14}$ Am$^{-2}$s$^{-1}$). The anstisymmetrisation process removes any contributions from extrinsic capacitance/impedance from the measurement circuit (which exhibits a constant value with $B$).
}
\label{fig_E6}
\end{figure}

\begin{figure}
\centering
\includegraphics[width=0.92\textwidth]{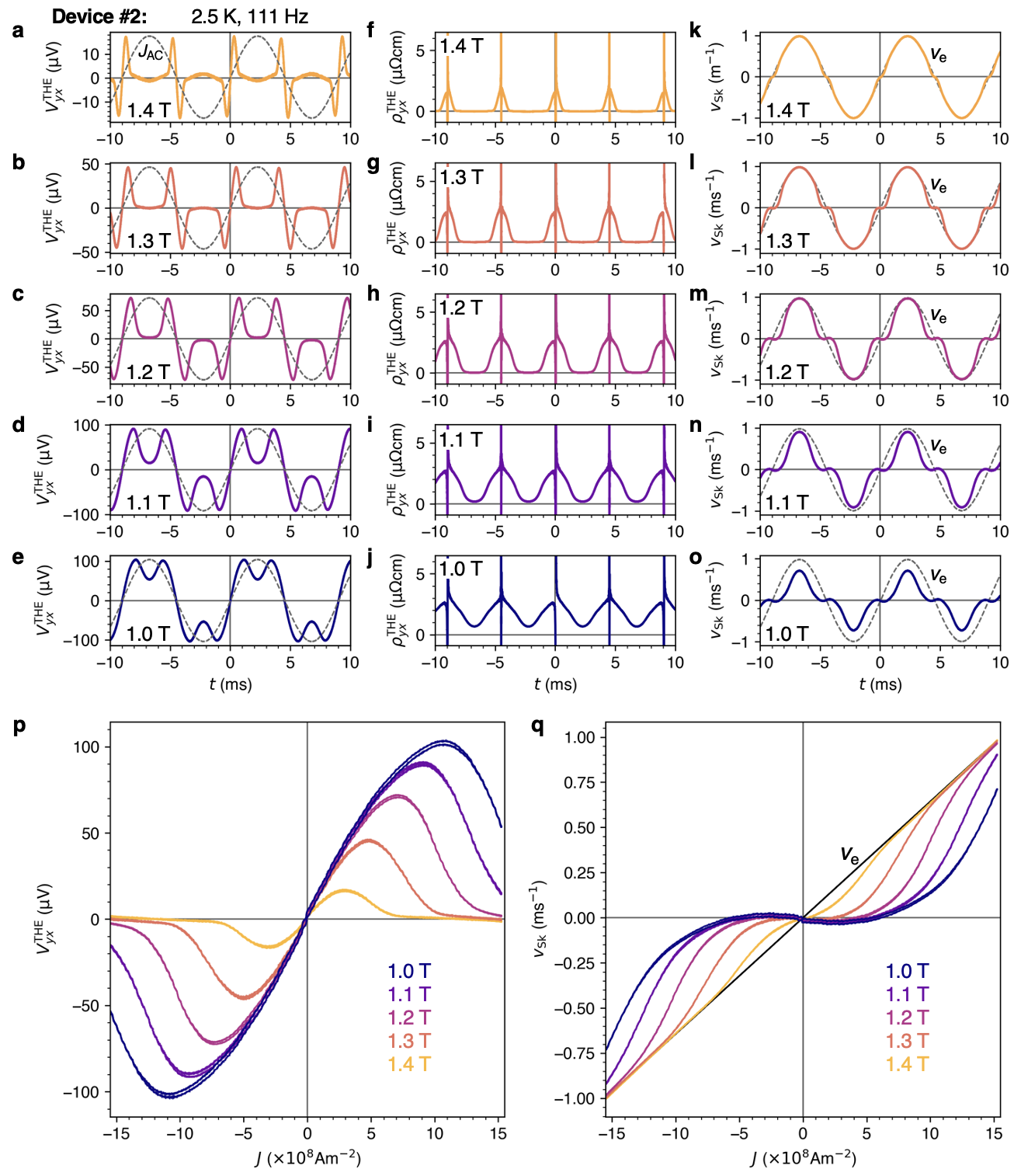}
\caption{$\vert$ \textbf{Field-dependent oscilloscope Hall voltage measurements.} \textbf{a}-\textbf{e} The antisymmetrised and isolated topological Hall voltage response $V_{yx}^{\rm{THE}}$ of Device \#2, measured at 2.5~K, and plotted as a function of time $t$. Data was acquired at a range of applied fields between 1.0 and 1.4~T, as shown by each panel respectively. The dashed grey line plots the corresponding applied sinusoidal current density $J$ with a peak $J_{\rm{AC}}^{\rm{max}}$ of 1.5$\times$10$^{9}$ Am$^{-2}$, and a frequency of 111 Hz. \textbf{f}-\textbf{j} The calculated topological Hall effect resistivity $\rho_{yx}^{\rm{THE}}$ at each field, plotted as a function of $t$. \textbf{k}-\textbf{o} The calculated skyrmion velocity $v_{\rm{Sk}}$ at each field. The electron drift velocity, $v_{e}$ calculated from the applied $J$, is plotted as the dashed grey line. \textbf{p}, \textbf{q} $V_{yx}^{\rm{THE}}$ and $v_{\rm{Sk}}$ plotted as a function of the applied $J$ for each applied magnetic field.
}
\label{fig_E7}
\end{figure}

\begin{figure}
\centering
\includegraphics[width=1\textwidth]{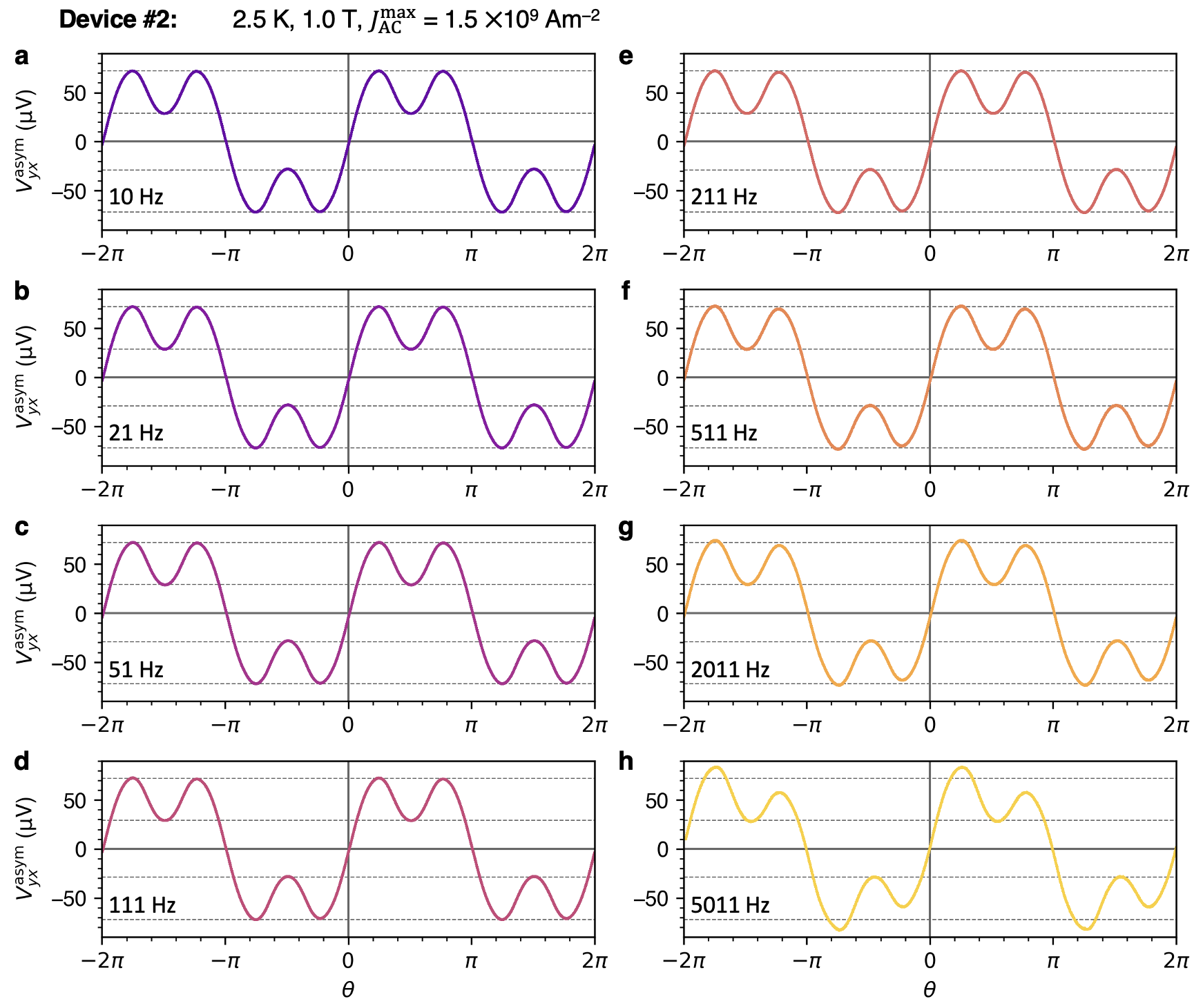}
\caption{$\vert$ \textbf{Frequency-dependent oscilloscope measurements of the Hall voltage.} \textbf{a}-\textbf{h} The antisymmetrised topological voltage response $V_{yx}^{\rm{THE}}$ of Device \#2, measured at 2.5~K and 1.0~T, plotted as a function of time $t$. Data was acquired with a sinusoidal current density (dashed grey line),with peak amplitude $J_{\rm{AC}}^{\rm{max}}$ of 1.5 $\times$ 10$^9$ Am$^{-2}$, and frequencies between 10 and 5011~Hz. The horizontal dashed lines highlight the inflection points of the voltage response at low frequencies. At high frequencies, the nonlinearity of the voltage response is delayed due to the inertial-like motion of the skyrmion lattice in the creep regime.
} 
\label{fig_E8}
\end{figure}

\begin{figure}
\centering
\includegraphics[width=1\textwidth]{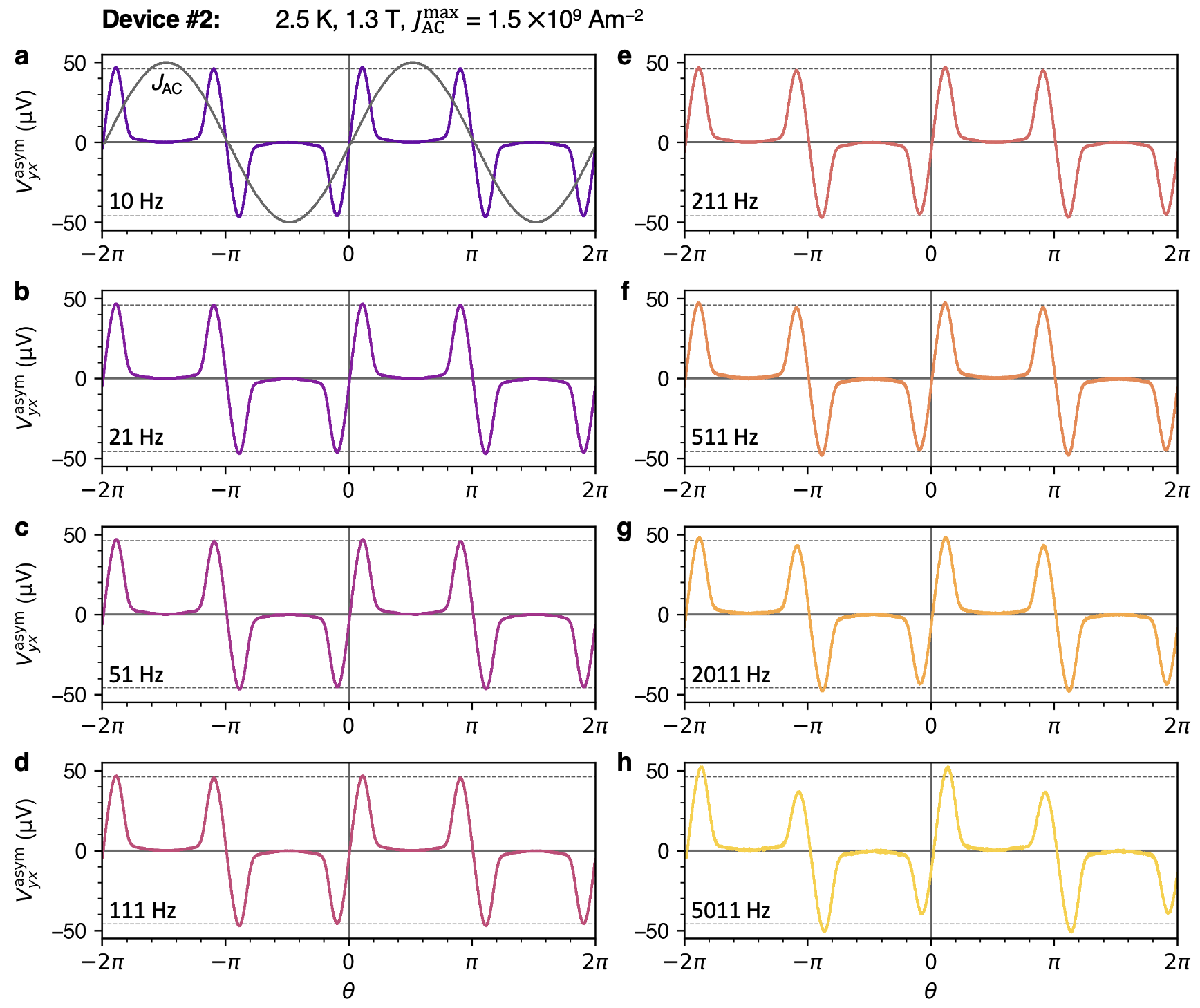}
\caption{$\vert$ \textbf{Frequency-dependent oscilloscope measurements of the Hall voltage.} \textbf{a}-\textbf{h} The antisymmetrised topological voltage response $V_{yx}^{\rm{THE}}$ of Device \#2, measured at 2.5~K and 1.3~T, plotted as a function of time $t$. Data was acquired with a sinusoidal current density (dashed grey line), with peak amplitude $J_{\rm{AC}}^{\rm{max}}$ of 1.5 $\times$ 10$^9$ Am$^{-2}$, and frequencies between 10 and 5011~Hz. The horizontal dashed lines highlight the inflection points of the voltage response at low frequencies. At high frequencies, the nonlinearity of the voltage response is delayed due to the inertial-like motion of the skyrmion lattice in the creep regime.
} 
\label{fig_E9}
\end{figure}

\begin{figure}
\centering
\includegraphics[width=0.75\textwidth]{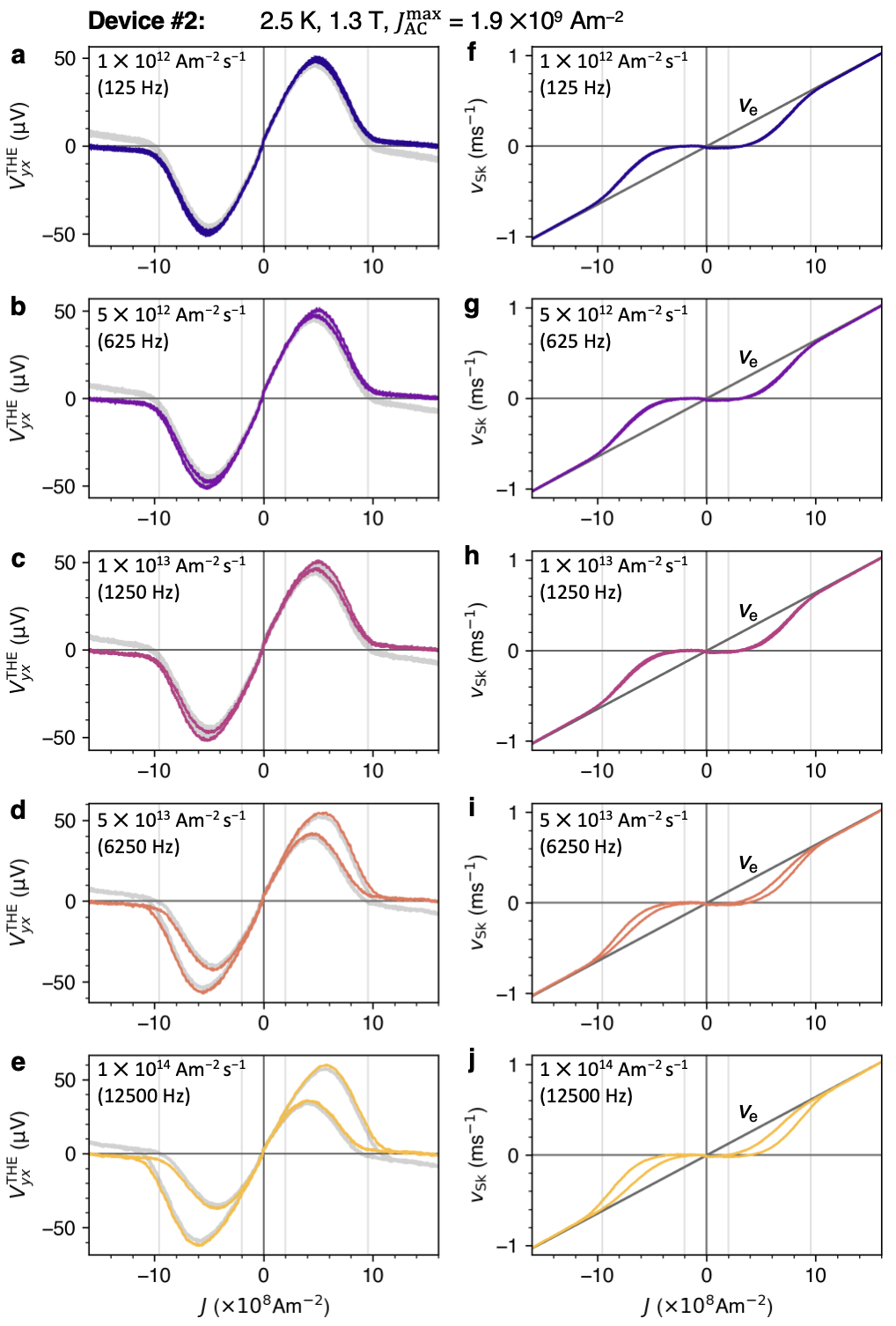}
\caption{$\vert$ \textbf{Frequency-dependent oscilloscope measurements of the Hall I-V characteristic and skyrmion velocity profile.} \textbf{a}-\textbf{e} The measured and isolated $V_{yx}^{\rm{THE}}$ measured for an applied oscillating triangle wave current density $J$ with peak amplitude $J_{\rm{AC}}^{\rm{max}}$ = 1.9 $\times$ 10$^{9}$ Am${-2}$, with current sweep rates $dJ/dt$ between 1 $\times$ 10$^{12}$ Am$^{-2}$s$^{-1}$ and 1 $\times$ 10$^{14}$ Am$^{-2}$s$^{-1}$. The corresponding frequencies are labelled. The grey lines plot the antisymmetrised voltage signal $V_{yx}^{\rm{asym}}$ before the ordinary Hall effect subtraction. \textbf{f}-\textbf{j} The calculated skyrmion velocities $v_{\rm{Sk}}$ plotted as a function of $J$ for each $dJ/dt$. The electron drift velocity $v_{\rm{e}}$ is plotted as the grey line.
} 
\label{fig_E10}
\end{figure}

\end{document}


\footnotesize
\maketitle
\begin{affiliations}
 \item RIKEN Center for Emergent Matter Science (CEMS), Wako, Saitama 351-0198, Japan.
 \item Department of Applied Physics, The University of Tokyo, Bunkyo-ku, Tokyo 113-8656, Japan.
 \item Tokyo College, University of Tokyo, Bunkyo-ku, Tokyo 113-8656, Japan.
\end{affiliations}

\newpage
\footnotesize

\section*{Supplementary Note 1: Derivation of skyrmion velocity formula}

Consider the experimental setup in Fig. S1. 
Here we study the simplest model for the conduction electrons described by the 
Hamiltonian as (with $\hbar=1$)
\begin{align}
H = \sum_{\bm{k}, \sigma}  \varepsilon_{\bm{k}} c^\dagger_{\bm{k} \sigma} c_{\bm{k} \sigma}
- J \sum_{i, \alpha, \beta} \bm{S}_i \bm{S}_i \cdot  c^\dagger_{i \alpha} \bm{\sigma}_{\alpha \beta} c_{i \beta},
\end{align}
where $\varepsilon_{\bm{k}} = \frac{\bm{k}
^2}{2m} - \mu$, and $\sigma, \alpha, \beta$ are spin component, and $\bm{\sigma}= (\sigma_x, \sigma_y, \sigma_z)$ are the Pauli matrices. Assume that the spin texture $\bm{S}_i$ is slowly varying, which is locally regarded as the ferromagnetic state, and redefine $\sigma= \pm 1$ as the spin component of the electrons parallel ($\sigma=1$) or antiparallel ($\sigma = -1$) to the spin texture. The formula for the topological Hall effect (THE) voltage $V_{yx}^{\rm{THE}}$ can be derived naively, by considering the two contributions from electrons with $\sigma =1$ and $\sigma = -1$ subject to the emergent electromagnetic force
due to a skyrmion with emergent magnetic field $\bf{b_{\rm{em}}}$ and with velocity $\bf{v_{\rm{Sk}}}$,
\begin{equation}
\bf{F} = -\rm{e\sigma }(\bf{v_{\rm{k}}} \times {b_{\rm{em}}}) +\rm{e \sigma}(\bf{v_{\rm{Sk}}} \times {b_{\rm{em}}}).
\label{eq:force}
\end{equation} 
The first term in eq.(\ref{eq:force}) gives rise to the Hall conductivity 
\begin{align}
\sigma_{xy} = \sigma_{xy}^{\sigma=1} + \sigma_{xy}^{\sigma=-1}
= P \frac{n e^2 \tau}{m} \omega_c \tau ,
\end{align}
with $P = \frac{\sum_\sigma \sigma n_{\sigma} }{n}$
where $n_{\sigma}$ is the electron density with $\sigma$ and $n = \sum_\sigma n_{\sigma}$ is the 
total electron density, $\omega_c = \frac{e b}{m}$ is the cyclotron frequency, and $\tau$ is the relaxation time 
assumed to be independent of $\sigma$.
On the other hand, the second term in eq.(\ref{eq:force}) leads to the
transverse current 
\begin{align}
J_{\rm tr}^{Sk} = - P \sigma_{xx} v_{\rm Sk} b_{\rm{em}},
\end{align}
with $\sigma_{xx} = \frac{ne^2 \tau}{m}$.
Therefore, the total transverse current is 
\begin{align}
J_{\rm tr} =  \sigma_{xy} E - P \sigma_{xx} v_{\rm Sk} b_{\rm{em}}.
\label{eq:jtr1}
\end{align}
Now the current in the direction of the electric field $E$ is given by $J = n e v_{\rm e}$
where $mv_{\rm e} = q$ with $q$ being the shift of the electron distribution in momentum space. 
Then $E= J/\sigma_{xx} = n e v_{\rm e}/(ne^2\tau/m) = m v_{\rm e}/(e \tau)$, and eq.(\ref{eq:jtr1}) becomes
\begin{align}
J_{\rm tr} = P \sigma_{xx} ( v_{\rm e} - v_{\rm Sk} ) b_{\rm{em}}.
\label{eq:jtr2}
\end{align}

\begin{figure}
\centering
\includegraphics[width=0.8\textwidth]{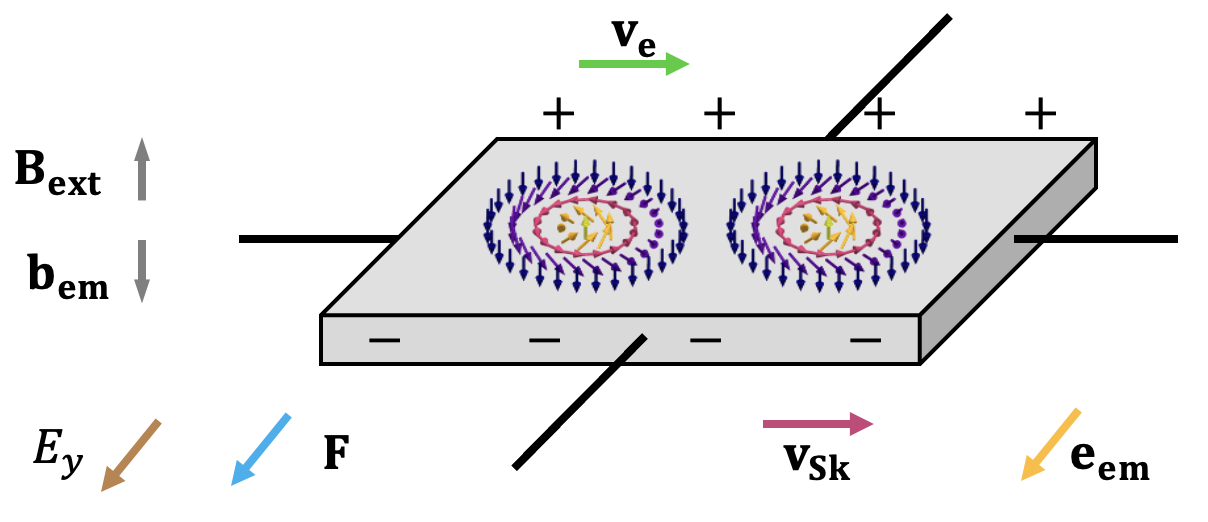}
\caption{
$\vert$ \textbf{Schematic of the Topological Hall effect voltage.} This diagram shows clearly the orienation of all vectors associated with the emergent electrodynamics of the SkL moving with velocity $\bf{v}_{\rm{Sk}}$ under flow of conduction electrons with velocity $\bf{v}_{\rm{e}}$. The relevant electric fields are the externally applied magnetic field $\bf{B}_{\rm{ext}}$, the emergent magnetic field $\bf{b}_{\rm{em}}$, and the emergent electric field induced by the SkL motion $\bf{e}_{\rm{em}}$. The emergent Lorentz forces on the electrons induced by the emergent magnetic and electric fields are $\bf{F}_{\rm{b}}$ and $\bf{F}_{\rm{e}}$, respectively. Finally, we label the resulting electric field setup by THE, $E_y$, which results in the measured THE voltage $V_{yx}^{\rm{THE}}$.
}
\label{figS1}
\end{figure}

The electric field $E_y$ along the transverse direction is determined by compensating
eq.(\ref{eq:jtr2}) in terms of $\sigma_{xx} E_y$, i.e., $E_y =  P ( v_{\rm e} - v_{\rm Sk} ) b_{\rm{em}}$.
Consequently the Hall voltage is
\begin{align}
V_{yx} =  wP b_{\rm{em}} (v_{\rm{e}} - v_{\rm{Sk}}),
\end{align}
reaching equation (2) in the main text, where $w$ is the width of the sample. Next, we substitute an expression for $\Delta V_{yx}^{\rm{THE}}=wJ\Delta\rho_{yx}^{\rm{THE}}(J)$ 
in terms of the current density $J$ and the change in the topological Hall resistivity $\Delta \rho_{yx}^{\rm{THE}}(J)$, 
and rearrange,
\begin{align}
v_{\rm{Sk}} = \frac{J\Delta\rho_{yx}(J)}{P b_{\rm{em}}}.
\end{align}

\section*{Supplementary Note 2: Galilean relativity in the multiband model}

We assume that the spin texture is slowly varying so that it can be regarded as almost ferromagnetic within the scale of lattice constant. In this case, we can use the Boltzmann transport theory for each band $n,\sigma$ where $n$ specifies the band and $\sigma$ the spin component parallel (+1) 
or anti-parallel ($-$1) to the magnetization at each spatial position $\bm{r}$. Therefore, the Boltzmann equation reads
\begin{align}
\frac{\partial f_{n \sigma}(\bm{k},\bm{r},t)}{\partial t} + \bm{v}_{n \sigma \bm{k}} \cdot \frac{\partial f_{n \sigma} 
(\bm{k},\bm{r},t)}{\partial \bm{r}}
+ \bm{F}_{\sigma} (\bm{r},t) \cdot  \frac{\partial f_{n \sigma} (\bm{k},\bm{r},t)}{\partial \bm{k}} = 
\biggl( \frac{\partial f_{n \sigma} (\bm{k},\bm{r},t)}{\partial t} \biggr)_{\rm coll.},
\label{eq:Bol}
\end{align}
where $\bm{v}_{n \sigma \bm{k}} = \frac{\partial \varepsilon_{n \sigma \bm{k}} }{\partial \bm{k} }$
is the group velocity of the conduction electron for each band, 
and the force $\bm{F}(\bm{r},t)$ is given by 
\begin{align} 
\bm{F}_{\sigma} (\bm{r},t) = -\sigma e ( - \bm{v} + \bm{v}_{n \sigma \bm{k}} ) \times \bm{b}( \bm{r} - \bm{v} t),
\end{align}
for the moving skyrmion lattice with the velocity $\bm{v}= \bm{v}_{\rm Sk}$ and with $\bm{b}(\bm{r})$ being the emergent magnetic field due to the skyrmion lattice. The right hand side of eq.(\ref{eq:Bol}) is the collision term which drives the system to the equilibrium state.

Now we define the moving frame by defining
$\bm{r}' = \bm{r} - \bm{v} t$ and $t' = t$, which leads to
\begin{align}
\frac{\partial}{\partial \bm{r}} &= \frac{\partial}{\partial \bm{r}'},   \nonumber \\
\frac{\partial}{\partial t} &= \frac{\partial}{\partial t'} - \bm{v} \cdot \frac{\partial}{\partial \bm{r}'}.
\label{eq:diff}
\end{align}
In this frame, the Boltzmann equation becomes
\begin{align}
\frac{\partial f'_{n \sigma}(\bm{k},\bm{r'},t')}{\partial t'} + (\bm{v}_{n \sigma \bm{k}} - \bm{v})  \cdot \frac{\partial f'_{n \sigma} 
(\bm{k},\bm{r}',t')}{\partial \bm{r}'}
-\sigma e ( - \bm{v} + \bm{v}_{n \sigma \bm{k}} ) \times \bm{b}( \bm{r}') \cdot  
\frac{\partial f'_{n \sigma} (\bm{k},\bm{r}',t')}{\partial \bm{k}} 
= \biggl( \frac{\partial f'_{n \sigma} (\bm{k},\bm{r}',t')}{\partial t'} \biggr)_{\rm coll.}.
\label{eq:Bol2}
\end{align}
Now we define $\bm{k}'_{n \sigma} = \bm{k} - \bm{q}_{n \sigma \bm{k}}$ for the moving frame by solving
\begin{align}
\bm{v}_{n \sigma \bm{k}' } =  \bm{v}_{n \sigma \bm{k}} - \bm{v}. 
\label{eq:v}
\end{align}
Considering the fact that the velocity $|\bm{v}|$ is much smaller than the Fermi velocity $v_F$,
it is enough to keep the linear order in $\bm{v}$, and 
eq.(\ref{eq:v}) can be solved by
\begin{align}
\sum_{\beta = x,y,z} (m^{-1}_{n \sigma})(\bm{k})_{\alpha, \beta} q_{n \sigma \bm{k} \beta} = v_{\alpha},
\end{align}
where
\begin{align}
(m_{n \sigma}^{-1}(\bm{k}))_{\alpha, \beta} = \frac{\partial^2 \varepsilon_{n \sigma \bm{k}} }{\partial k_\alpha \partial k_\beta}
\end{align}
is the inverse of the mass tensor. Here we assume that the mass of each band is constant, i.e., 
$(m_{n \sigma}^{-1}(\bm{k}))_{\alpha, \beta} = m_{n \sigma}^{-1} \delta_{\alpha, \beta}$.
Then eq.(\ref{eq:Bol2}) becomes
\begin{align}
\frac{\partial f'_{n \sigma}(\bm{k}',\bm{r}',t')}{\partial t'} + \bm{v}_{n \sigma \bm{k}'}  
\cdot \frac{\partial f'_{n \sigma} (\bm{k}',\bm{r}',t')}{\partial \bm{r}'}
-\sigma e  \bm{v}_{n \bm{k}'}  \times \bm{b}( \bm{r}') \cdot  \frac{\partial f'_{n \sigma} (\bm{k}',\bm{r}',t')}{\partial \bm{k}'} 
= \biggl( \frac{\partial f'_{n \sigma} (\bm{k}',\bm{r}',t')}{\partial t'} \biggr)_{\rm coll.}.
\label{eq:Bol3}
\end{align}
The left hand side of this Boltzmann equation is that for the equilibrium state under the static
emergent magnetic field. Now the assumption is that the collision term on the right hand side
relaxes the system to the equilibrium in the moving frame. Namely, the system adjusts itself searching for the moving frame to minimize the energy. This is analogous to the original idea of Fr\"olich superconductivity in the density wave state\cite{frohlich_theory_1954,lee_conductivity_1974},
where the free energy is minimized in the moving frame for the sliding density wave state.
The total cancellation of the Hall voltage observed experimentally implies that the 
mechanism to determine the shift of the Fermi surface for each band is to minimize the 
energy instead of the balance between the acceleration by the electric field and the
relaxation.  

We can estimate the condition when this assumption is justified as follows (below we drop the band index $n$ and spin index $\sigma$, and $\hbar =  1$). The shift $q$ is typically given by $m v$ with $m$ being the mass of the electrons. Then the right hand side of eq.(\ref{eq:Bol3}) is estimated as $\sim \frac{m v v_k}{\tau} \frac{\partial f}{\partial \varepsilon}$, where $\tau$ is the 
relaxation time. This should be compared with the perturbation due to $v$ on the left side of 
eq.(\ref{eq:Bol3}), i.e., drift terms.
The second term is estimated as 
$\sim v \frac{\partial \varepsilon}{\partial r} \frac{\partial f}{\partial \varepsilon}$,
while the third term 
$\sim v eb v_k \frac{\partial \varepsilon}{\partial k} \frac{\partial f}{\partial \varepsilon}$.
Consider that
$\frac{\partial \varepsilon}{\partial r} \sim \frac{\varepsilon}{\lambda}$, where $\lambda$ is the skyrmion size,
and $v_k = \frac{\partial \varepsilon}{\partial r} \sim v_F \sim 
\frac{\varepsilon}{k_F} \sim a \varepsilon$
where $v_F$ is the Fermi velocity, $k_F$ is the Fermi wavenumber, and
$a \sim 1/k_F$ is the lattice constant. Moreover, the emergent magnetic field $e b \sim 1/\lambda^2$. 
Putting these together, the ratio is estimated as
\begin{align}
\frac{\rm r.h.s. \ \ of \ \ eq.(\ref{eq:Bol3}) }{\rm second \ \ term \ \ on \ \  l.h.s \ \ of \ \ eq.(\ref{eq:Bol3}) } 
\sim \frac{\lambda}{\ell},
\label{eq:rel1}
\end{align}
and 
\begin{align}
\frac{\rm r.h.s. \ \  of \ \ eq.(\ref{eq:Bol3}) }{\rm third \ \ term \ \ on \ \ l.h.s \ \ of \ \ eq.(\ref{eq:Bol3} )} &\sim 
\biggl( \frac{\lambda}{a} \biggr)^2 \frac{1}{k_F \ell}.  
\label{eq:rel2}
\end{align}
In our system, $\lambda$ is 2.5~nm, $a$ is roughly 0.4~nm, 
such that $\lambda/a \sim 6$, while $k_F \ell \sim \ell/a $ is estimated to be $\sim 14$. 
In this scenario, the condition for the Galilean relativity is much more relaxed compared with that
derived in the past work\cite{schulz_emergent_2012}, and is applicable to the present material.
Therefore, the ratio in (\ref{eq:rel1}) is $\sim 0.4 <1$ while that in (\ref{eq:rel1}) is $\sim 2.5 >1$.
Therefore, the shift $q$ is mostly determined by the energy minimization in the comoving frame, while
the approximation $\omega_c \tau <1$ is justified.
 
\section*{Supplementary Note 3: Two-band model}
To clarify the meaning of the emergent Galilean relativity, we applied to the formalism to the two band model, i.e., the two spin component model treated in Supplementary Note 1, where the relaxation time $\tau$ and the mass $m$ were assumed to be identical. Here, we consider the genetic case where $m_{\sigma = +1} \ne m_{\sigma = -1}$ and $\tau_{\sigma = +1} \ne \tau_{\sigma = -1}$. Then we can define $\sigma_{xx}^{\sigma = +1}$ and $\sigma_{xx}^{\sigma = -1}$ accordingly. After some calculations, one obtains that
\begin{align}
J_{\rm tr} = \sigma_{xx} ( P_e v_{\rm e} - P_{\rm sk} v_{\rm Sk} ) b_{\rm{em}},
\end{align}
with the coefficients $P_e = \frac{1}{2} \biggl[ \frac{ (1+ P)^2}{1+Q} - \frac{ (1-P)^2}{1-Q} \biggr]$, and
$P_{\rm sk} = Q$ with
$P = \frac{ \sigma_{xx}^{\sigma = +1} -  \sigma_{xx}^{\sigma = -1}}{\sigma_{xx}^{\sigma = +1} +  \sigma_{xx}^{\sigma = -1}}$
and $Q= \frac{ n^{\sigma = +1} -  n^{\sigma = -1} }{ n^{\sigma = +1} + n^{\sigma = -1} }$.
In general, $ P_e \ne P_{\rm sk}$.  Also $v_{\rm Sk}$ as a function of $v_{\rm e}$ should be determined by
the Thiele equation, where the spin transfer torque is driven by the spin current, not the charge current. Therefore, the relation $v_{\rm e} = v_{\rm Sk}$ is highly nontrivial in this formulation.

On the other hand, the emergent Galilean picture in the clean limit, one can start with the more general and microscopic model with the Lagrangian:
\begin{align}
L = \int d x \biggl[ i \psi_\sigma (x,t) \partial_t \psi_\sigma (x,t) + 
\psi_\sigma (x,t)  \biggl(- \frac{ \partial_x^2}{m_\sigma} + \mu \biggr) \psi_\sigma (x,t)
+ \psi^\dagger_\sigma (x,t) V_{\sigma, \sigma'} (x - v_{\rm sk} t) \psi_{\sigma'} (x,t)  \biggr],
\end{align}
where we put $\hbar=1$.
Note that the skyrmion crystal velocity  $v_{\rm sk}$ as a function of the current $J_{\rm ext.}$ is determined 
self-consistently as follows. Assume $v_{\rm sk}$ as given above, the electrons motion is induced to 
result in the current $J$. Then this current $J$ should be equal to the external current $J_{\rm ext.}$ 
to drive the skyrmions.
By the Galilean transformation 
\begin{align}
x' &= x - v_{\rm sk} t   \nonumber  \\
t' &= t   \nonumber \\ 
\psi'_\sigma &= e^{- i \eta_\sigma (x)} \psi_\sigma \nonumber \\ 
\psi'^\dagger_\sigma &= e^{ i \eta_\sigma (x)} \psi^\dagger_\sigma, 
\label{eq:GT}
\end{align}
with $\eta_\sigma(x) = m_\sigma v_{\rm sk}$, 
one obtains 
\begin{align}
L' = \int d x' \biggl[ i \psi'_\sigma (x',t') \partial_{t'} \psi'_\sigma (x',t') + 
\psi'_\sigma (x',t')  \biggl(- \frac{ \partial_x'^2}{m_\sigma} + \mu - \frac{m_\sigma}{2} v_{\rm sk}^2 \biggr) \psi'_\sigma (x',t')
+ \psi'^\dagger_\sigma (x',t') V_{\sigma, \sigma'} (x') \psi'_{\sigma'} (x',t') \biggr] 
\end{align}
for the comoving frame.
The principle is that the lowest energy state is realized in the frame where the 
Lagrangian or Hamiltonian is time-independent, i.e., static. Therefore, the ground state in the comoving frame is realized.  It is obvious that the momentum shift $q_\sigma$ of each band $\sigma$ is $m_\sigma v_{\rm sk}$ from eq.(\ref{eq:GT}), the contribution to the current from the band $\sigma$ is 
\begin{align}
J_\sigma &= \int \frac{d^d k}{(2 \pi)^2} - e \frac{k}{m_\sigma} f ( \varepsilon_\sigma(k-q_\sigma) ) \nonumber \\
&=  \int \frac{d^d k'}{(2 \pi)^2} - e \frac{k'+q_\sigma }{m_\sigma} f ( \varepsilon_\sigma(k') ) \nonumber \\
&= - e \frac{n^\sigma q_\sigma }{m_\sigma}.
\end{align}

\begin{figure}
\centering
\includegraphics[width=0.9\textwidth]{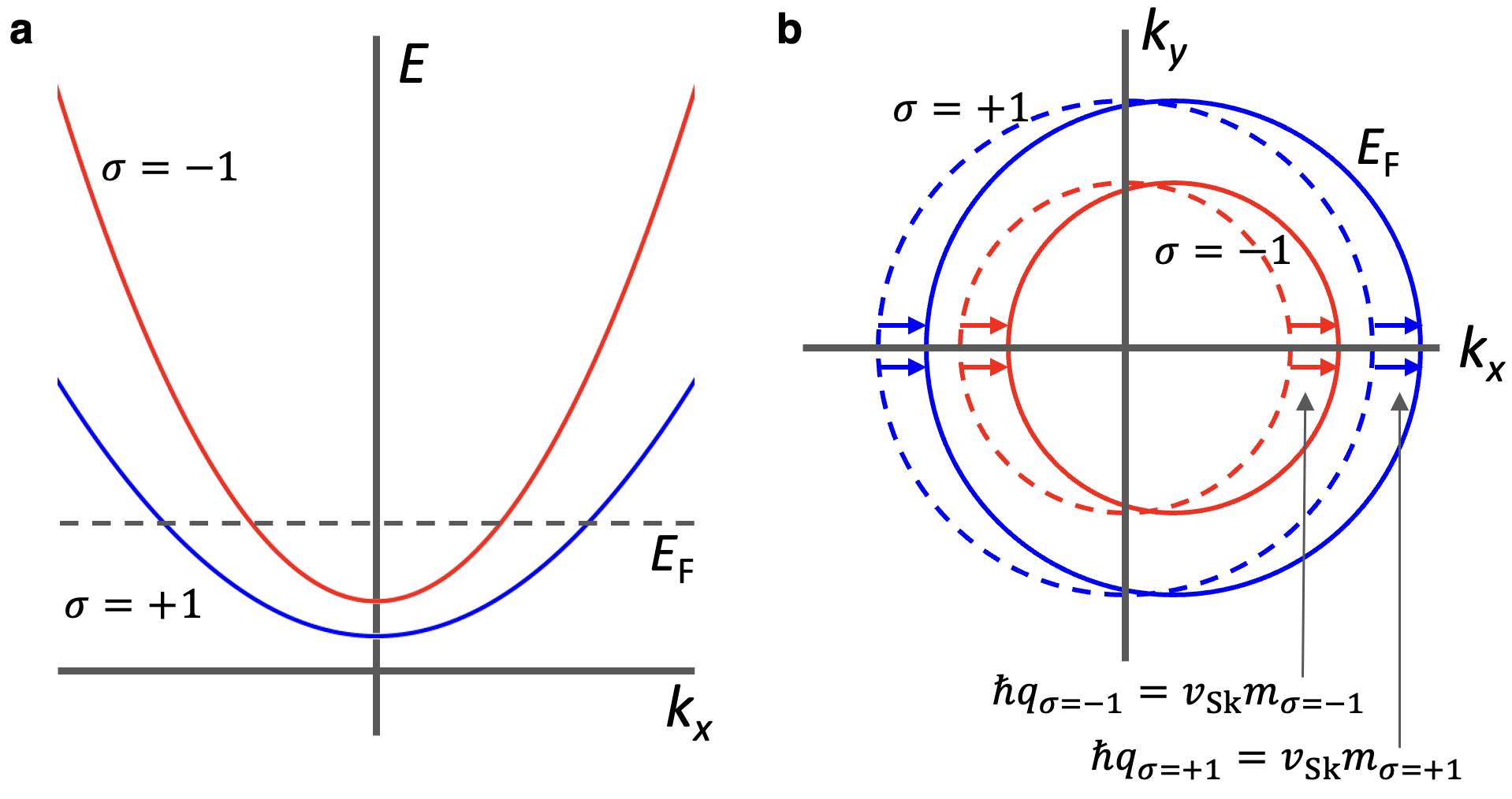}
\caption{
$\vert$ \textbf{Schematic illustration of the electron momentum shift in the two band model.} \textbf{a} A schematic illustration of the two band model, with two spin components $\sigma = \pm 1$, where $m_{\sigma = +1} \ne m_{\sigma = -1}$, and where $E_{\rm{F}}$ denotes the Fermi surface. \textbf{b}, A schematic illustration of the shift in the electron momentum distribution, $q_{\sigma = \pm 1}$. In the clean limit, this is defined by the effective mass of each band, and the skyrmion velocity $v_{\rm{Sk}}$.
}
\label{figS2}
\end{figure}

Then we obtain the total current in the presence of the moving skyrmion crystal as
$J = \sum_\sigma J_\sigma = - e \sum_\sigma \frac{n^\sigma q_\sigma }{m_\sigma} =  
- e \sum_\sigma n^\sigma v_{\rm sk} = - e n v_{\rm sk}$
where $n =\sum_\sigma n^\sigma$ is the total electron density.
This gives the electron velocity $v_e$ from the relation $J = - en v_e$, and we 
obtain the result $v_{\rm sk}= v_e$ in the flow region of the clean sample. The ideas behind this two band model are shown schematically in Fig. S2, which displays a simple two band model with two parabolic bands, and shows the corresponding shift $q_{\sigma = \pm 1}$ of the Fermi surface of the two bands, which is primarily determined by $v_{\rm sk}$ and $m_{\sigma = \pm 1}$ in the clean limit.

\section*{Supplementary Note 4: Joule heating considerations}
Here, we consider the possible effects of Joule heating in our measurements, and determine that they should not be significant. Firstly, we estimate the power generated due to the resistive heating of the sample: our samples showed a typical two terminal resistance of 20 $\Omega$, while the maximum current applied was 3~mA. This corresponds to a power $P=RI^2$ of roughly 0.2 mW. For a similar device composed of an FIB-fabricated lamella MnSi and a CaF$_2$ substrate, the authors of a previous work estimated a temperature change of $\sim$0.2~K for a similar dissipative power of 0.23 mW\cite{sato_nonthermal_2022}. Thus, we argue that this heating power should not be sufficient to raise the temperature of our device by more than a few degrees at most, while around 15~K heating would be required for the sample to exit the skyrmion phase entirely.

Further evidence for the current-induced motion origin of our results lies in the frequency dependence of the nonlinear Hall signal. In particular, due to the timescale of thermal processes (milliseconds), and thus delay of the system to reach a steady state following the application of a current, we would expect any heating effect should depend strongly on the frequency of the applied AC current. At high frequencies, the temperature would no longer be able to react the oscillating current, and would instead tend to some constant value. Similarly, this would lead to a constant suppression of the THE in the Joule heating scenario. However, in Extended Data Fig. 6-8, the suppression of the Hall voltage is the same for a range of measurements with sinusoidal currents applied at between 10 and 12500 Hz (and also in the DC limit, by comparison to Fig. 2 and 3, and Extended Data Fig. 2 and 3). Thus, we do not see indication of the system tending towards a constant temperature, which would be expected for a Joule heating scenario. Finally, we consider the presence of the current density thresholds for the onset of creep and flow motion -- $J_{\rm{th}}^{\rm{C}}$ and $J_{\rm{th}}^{\rm{F}}$, respectively. Particularly for $J_{\rm{th}}^{\rm{C}}$, it is difficult to see a scenario where the Joule heating exhibits some threshold current, followed by the linear scaling of $\rho_{yx}$, as shown in Fig. 2. Instead, we argue this is more naturally explained by the dynamics transition of the current-induced motion of the SkL state.

\begin{figure}
\centering
\includegraphics[width=0.8\textwidth]{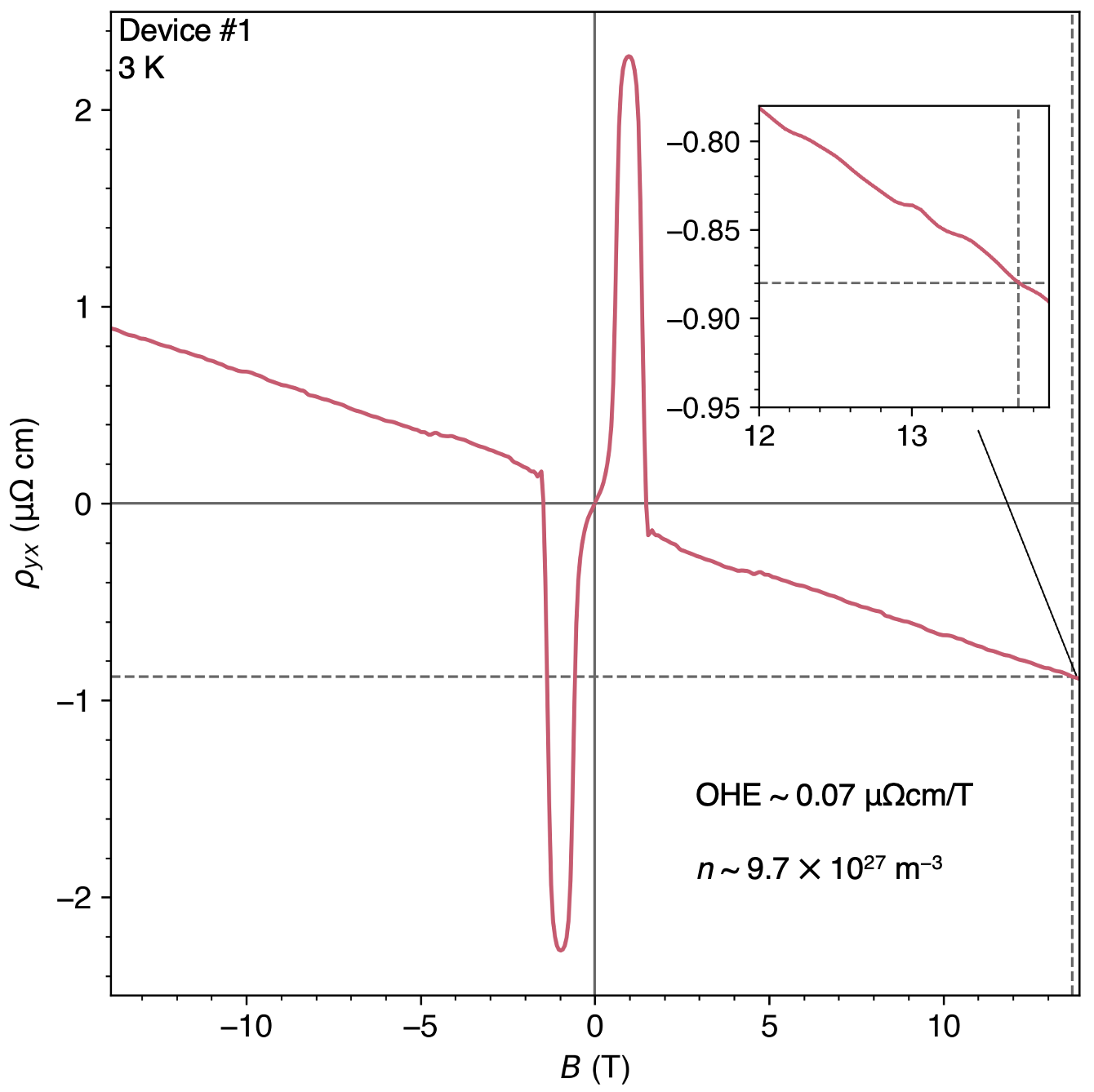}
\caption{$\vert$ \textbf{Measurement of the ordinary Hall effect and determination of the charge carrier density.} The Hall resistivity $\rho_{yx}$ of Gd$_2$PdSi$_3$ Device \#1 is plotted as a function of the applied magnetic field $B$, measured at 3~K. At high fields, above 10~T, we assume that the measured Hall signal is dominated by the ordinary Hall effect (OHE), as indicated by the linear dependence above 5~T. We utilise the measured value to estimate that the OHE contributes a value of $-$0.07 \textmu$\Omega$cm$^{-1}$ to $\rho_{yx}$. This value was subtracted from all measurements to isolate the topological Hall effect from the skyrmion phase. Oscilloscope measurements of the OHE at 9~T in Supplementary Figure S5 demonstrate this to be a reasonable approximation. The measured Hall voltage $V_\mathrm{H}$ at 13.7~T is used to estimate the charge carrier density in Gd$_2$PdSi$_3$, $n=BI/etV_H$, where $e$ is the electron charge, $I$ is the applied current (2~mA) and $t$ is the thickness of the sample (850~nm). The acquired value of 9.7$\times$10$^{27}$ m$^{-3}$ was used to calculate the electron drift velocity for a given current density $J$ using the standard Drude model formula, $v_{d}=J/ne$.
}
\label{fig_S3}
\end{figure}

\begin{figure}
\centering
\includegraphics[width=1\textwidth]{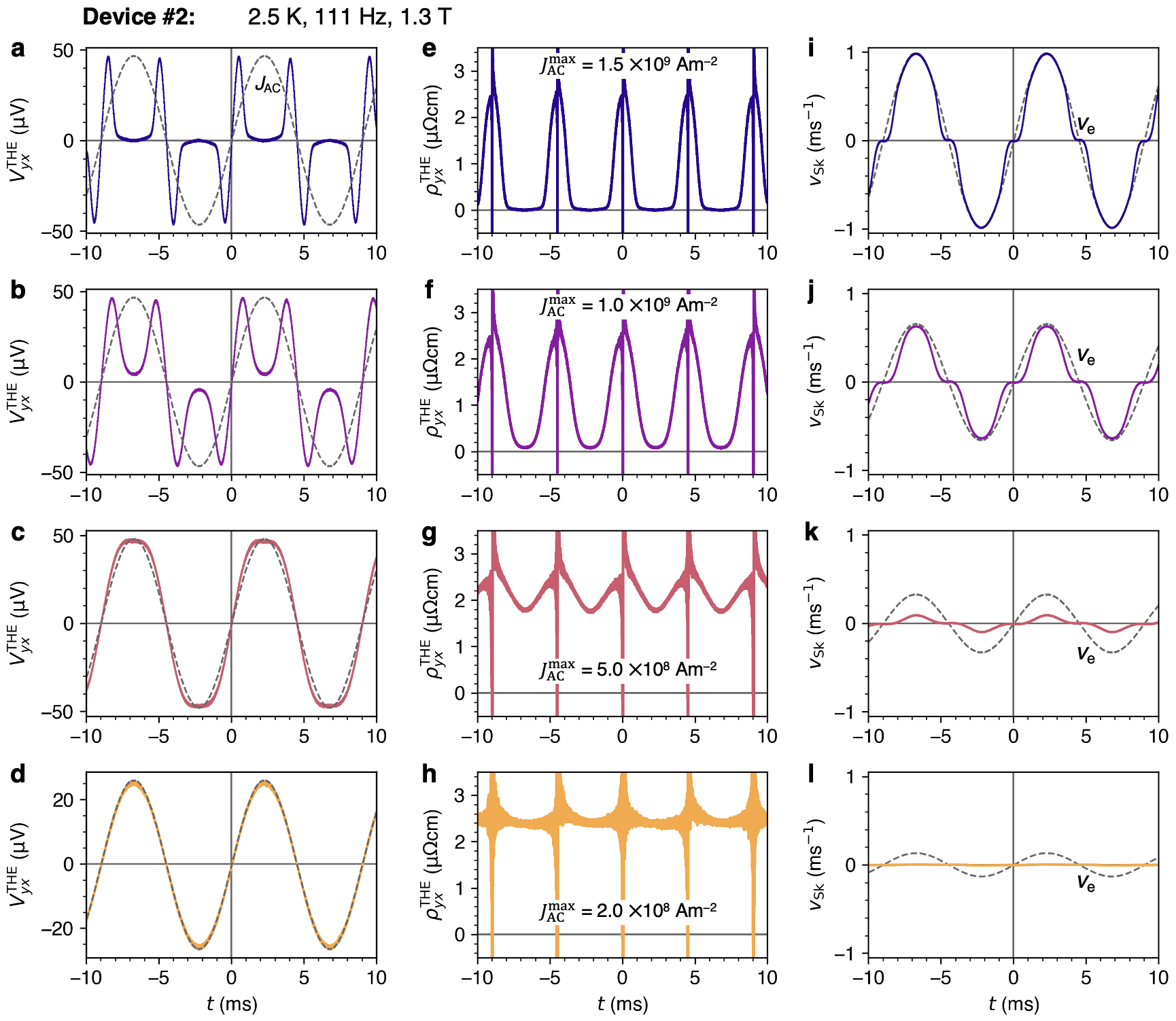}
\caption{$\vert$ \textbf{Current-dependent oscilloscope measurements of the topological Hall voltage.} \textbf{a}-\textbf{d} The antisymmetrised and isolated topological Hall voltage response $V_{yx}^{\rm{THE}}$ of Device \#2, measured at 2.5~K and 1.3~T, plotted as a function of time $t$. Data was acquired with a sinusoidal current density $J_{\rm{AC}}$ (dashed grey line), with varied peak amplitude $J_{\rm{AC}}^{\rm{max}}$ between 2.0$\times$10$^{8}$ Am$^{-2}$ and 1.5$\times$10$^{9}$ Am$^{-2}$, as labelled, and a frequency of 111~Hz. \textbf{e}-\textbf{h} The calculated topological Hall effect resistivity $\rho_{yx}^{\rm{THE}}$ for each applied current density. \textbf{i}-\textbf{l} The calculated skyrmion velocity $v_{\rm{Sk}}$ at each $J_{\rm{AC}}^{\rm{max}}$. The electron drift velocity, $v_{e}$ calculated from the applied $J$, is plotted as the dashed grey line.
} 
\label{fig_S4}
\end{figure}

\begin{figure}
\centering
\includegraphics[width=0.95\textwidth]{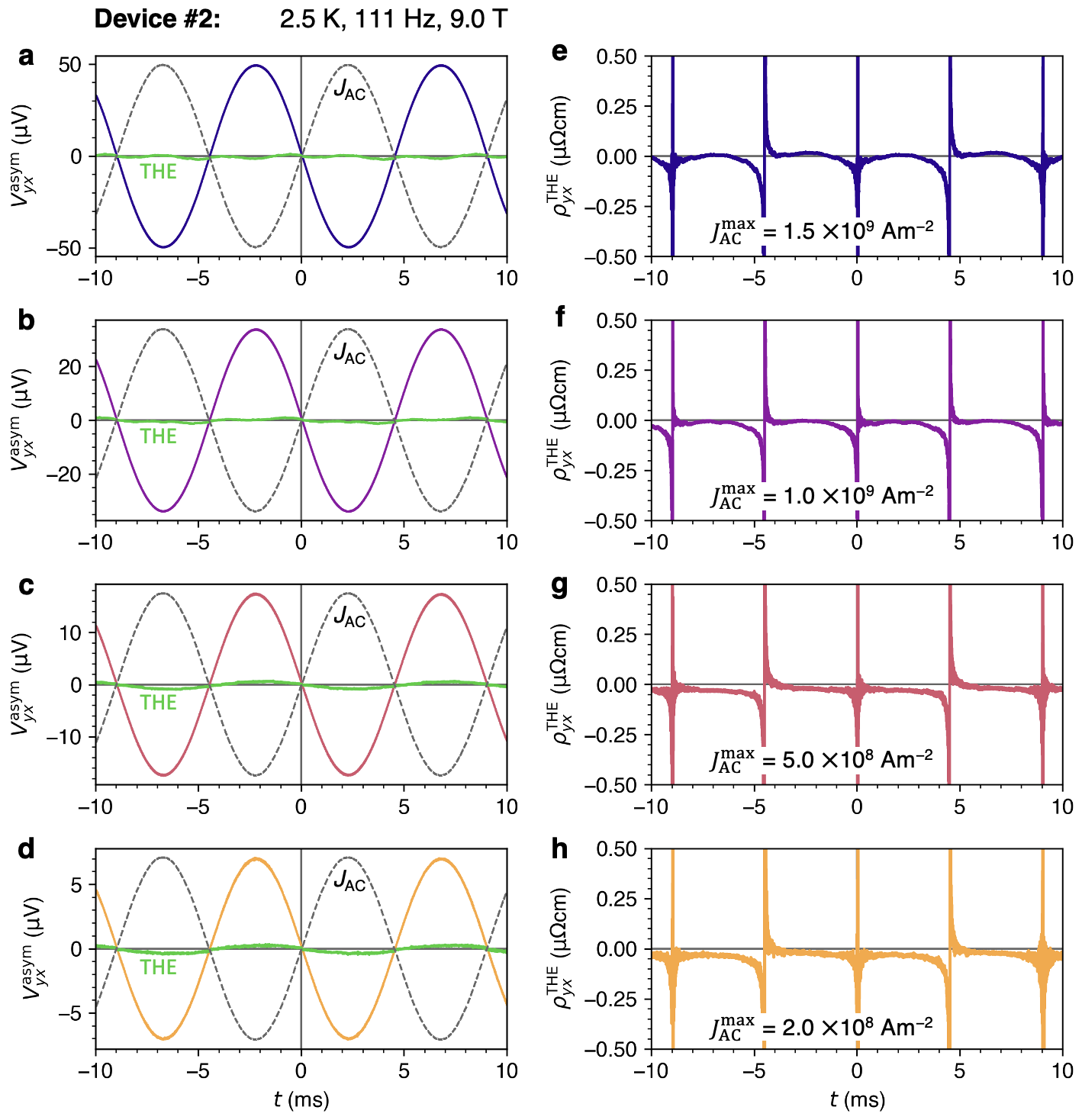}
\caption{$\vert$ \textbf{Current-dependent oscilloscope measurements of the ordinary Hall effect voltage.} \textbf{a}-\textbf{d} The antisymmetrised voltage response $V_{yx}^{\rm{asym}}$ of Device \#2, measured at 2.5~K and 9.0~T, plotted as a function of time $t$. Data was acquired with a sinusoidal current density (dashed grey line), with varied peak amplitude $J_{\rm{AC}}^{\rm{max}}$ between 2.0$\times$10$^{8}$ Am$^{-2}$ and 1.5$\times$10$^{9}$ Am$^{-2}$, as labelled, at a frequency of 111~Hz. The result of subtracting the measured Hall voltage and the estimated OHE value (utilised to determine the topological Hall effect (THE) contribution) is plotted as the green line. \textbf{e}-\textbf{h} The calculated topological Hall effect resistivity $\rho_{yx}^{\rm{THE}}$ for each applied current density, showing the absence of the THE and the lack of significant Hall effect nonlinearity outside of the SkL phase.
} 
\label{fig_S5}
\end{figure}

\begin{figure}
\centering
\includegraphics[width=0.78\textwidth]{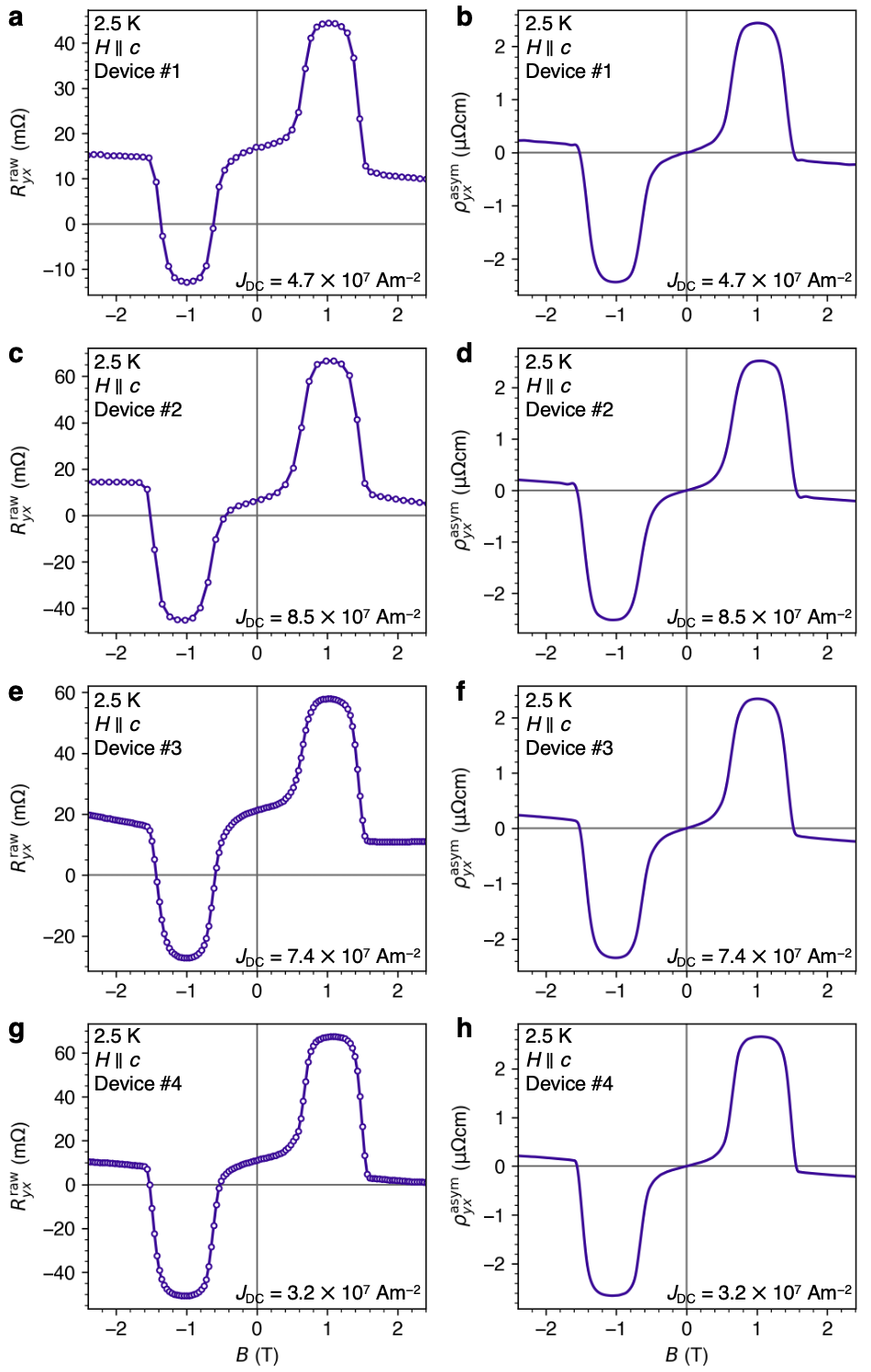}
\caption{$\vert$ \textbf{Comparison of raw and antisymmetrised Hall data for each device.} \textbf{a} The measured raw Hall resistance $R_{yx}^{\rm{raw}}$ as a function of the applied field $B$ at 2.5 K in Device \#1. \textbf{b} The corresponding antisymmetrised Hall resistivity $\rho_{yx}^{\rm{asym}}$. \textbf{c}-\textbf{h} The same, but for Devices \#2, \#3 and \#4.
} 
\label{fig_S6}
\end{figure}

\begin{figure}
\centering
\includegraphics[width=1\textwidth]{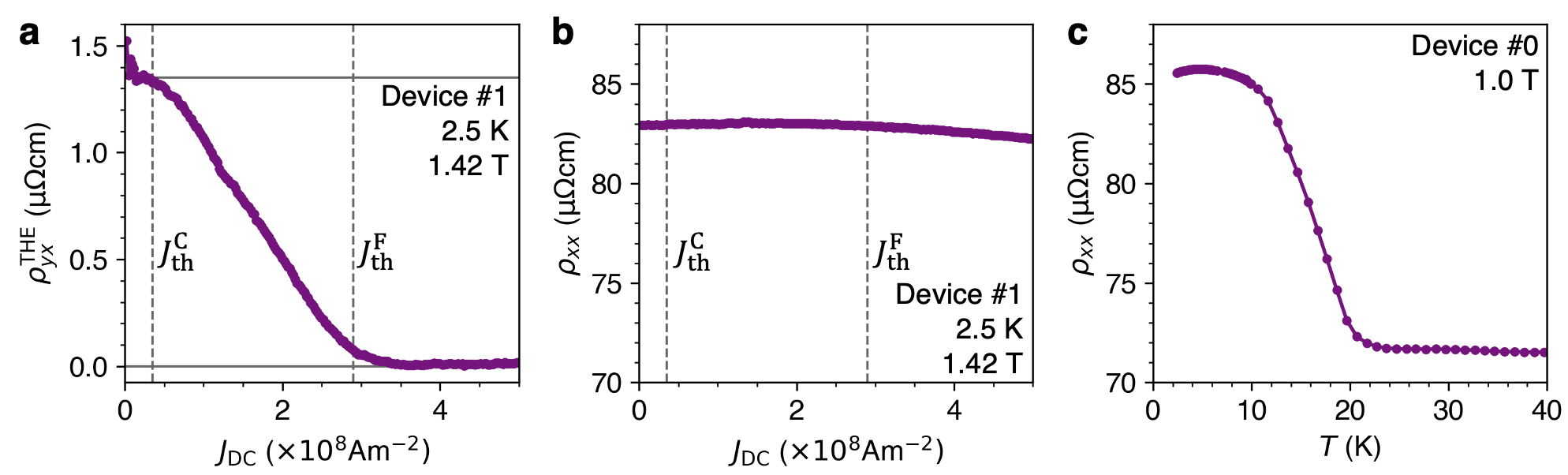}
\caption{$\vert$ \textbf{Comparison of current dependence of the Hall and longitudinal resistivity.} \textbf{a} The measured topological Hall effect resistivity, $\rho_{yx}^{\rm{THE}}$, as a function of the current density $J_{\rm{DC}}$, reproduced from Fig. 2 of the main text. \textbf{b} The corresponding longitudinal resistivity, $\rho_{xx}$, as a function of $J_{\rm{DC}}$. \textbf{c} The value of $\rho_{xx}$ measured as a function of $T$ for comparison, reproduced from Extended Data Fig. 1.
} 
\label{fig_S7}
\end{figure}

\begin{figure}
\centering
\includegraphics[width=0.8\textwidth]{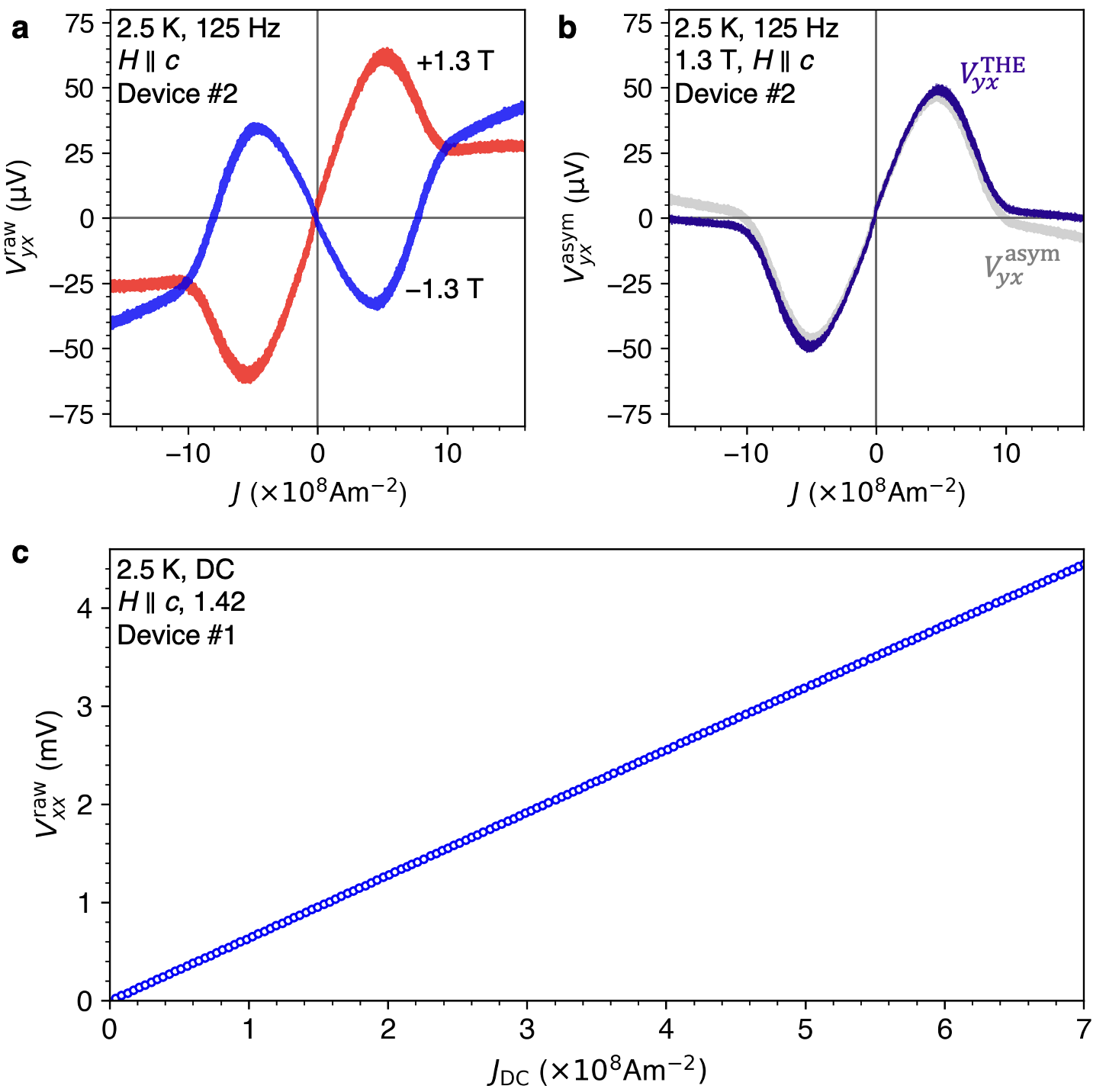}
\caption{$\vert$ \textbf{Raw voltage-current curve examples.} \textbf{a} The raw Hall voltage $V_yx^{\rm{raw}}$ measured as a function of $J$ in Device \#2 at 2.5, under an applied field of $\pm$1.3 T. This is equivalent to the data in Extended Data Fig. 6a. \textbf{b} The antisymmetrised Hall voltage $V_yx^{\rm{asym}}$, and the topological Hall voltage $V_yx^{\rm{raw}}$ (after subtracting the ordinary Hall effect component), plotted as a function of $J$, calculated from the data in panel \textbf{a}. \textbf{c} An example of the raw longitudinal voltage $V_xx^{\rm{raw}}$ measured as a function of the current density $J_{\rm{DC}}$ in Device \#1, showing nearly linear, Ohmic behavior (the same data as shown in Supplementary Fig. S7b).
} 
\label{fig_S8}
\end{figure}